\documentclass[aps,12pt,preprint,tightenlines,floatfix]{revtex4}

\newcommand{\figwl}{0.8}

\usepackage{amsmath}
\usepackage{amsfonts}
\usepackage{graphicx}

\newcommand{\EXP}[1]{\exp\left\{#1\right\}}
\newcommand{\average}[1]{\langle #1\rangle}
\newcommand{\laverage}[1]{\big\langle #1\big\rangle}
\newcommand{\identity}{\mathrm{1}\hspace{-0.53ex}\mathrm{l}}
\newcommand{\eq}[1]{(\ref{#1})}
\newcommand{\Eq}[1]{Eq.~\eq{#1}}
\newcommand{\Eqs}[2]{Eqs.~\eq{#1} and \eq{#2}}
\DeclareMathOperator{\erf}{erf}

\begin{document}

\title{Stationary and Transient Work-Fluctuation Theorems for a Dragged
Brownian Particle}
\author{R. van Zon and E.G.D. Cohen}
\affiliation{The Rockefeller University,
1230 York Avenue, New York, New York 10021}

\date{December 12, 2002}

\begin{abstract}
Recently Wang {\em et al.\/} carried out a laboratory experiment,
where a Brownian particle was dragged through a fluid by a harmonic
force with constant velocity of its center.  This experiment confirmed
a theoretically predicted work related integrated~(I) Transient
Fluctuation Theorem (ITFT), which gives an expression for the ratio
for the probability to find positive or negative values for the
fluctuations of the total work done on the system in a given time in a
transient state.  The corresponding integrated stationary state
fluctuation theorem (ISSFT) was not observed.  Using an overdamped
Langevin equation and an arbitrary motion for the center of the
harmonic force, all quantities of interest for these theorems and the
corresponding non-integrated ones (TFT and SSFT, resp.) are
theoretically explicitly obtained in this paper.  While the (I)TFT is
satisfied for all times, the (I)SSFT only holds asymptotically in
time.  Suggestions for further experiments with arbitrary velocity of
the harmonic force and in which also the ISSFT could be observed, are
given.  In addition, a non-trivial long-time relation between the ITFT
and the ISSFT was discovered, which could be observed experimentally,
especially in the case of a resonant circular motion of the center of
the harmonic force.
\end{abstract}

\maketitle

\section{Introduction}

Fluctuations of physical properties of statistical mechanical systems
were first considered, in the modern context of dynamical Hamiltonian
or dissipative systems theory, by Evans, Cohen and
Morriss\cite{Evansetal93}. It concerned here the statistics of phase
space contraction or entropy production fluctuations over a certain
time interval. In particular, the probabilities for equal positive or
negative entropy production fluctuations of a certain magnitude were
considered.  Two different physical situations have been
treated. First, in Ref.~\cite{Evansetal93}, for a non-equilibrium
stationary state, possibly far from equilibrium, the fluctuations of
the dissipative (viscous) part of the pressure tensor of a fluid were
studied. Next, Evans and Searles\cite{Evansetal94} studied the
fluctuations of entropy production in an ensemble of phase space
trajectories emanating from an initial equilibrium state in the course
of time.  While the first case concerned a study of stationary state
fluctuations in trajectory segments of a given duration along a single
trajectory in a non-equilibrium stationary state and will be called
the Stationary State Fluctuation Theorem (SSFT), the second case
involved a study of an ensemble of many transient phase space
trajectories each over a time $\tau$, all emanating from an
equilibrium ensemble at time $t=0$, which will be called a Transient
Fluctuation Theorem (TFT).

Mathematical proofs have been given of both
theorems\cite{Evansetal94,GallavottiCohen95a,GallavottiCohen95b,SearlesEvans00}
and many computer simulations have confirmed both
theorems(e.g. \cite{Evansetal93,Evansetal94,SearlesEvans00}).  While
the original proofs of both FT's were based on the deterministic
dynamics of many particles, later proofs for systems with stochastic
dynamics were given by Kurchan\cite{Kurchan98} and Lebowitz and
Spohn\cite{LebowitzSpohn99}. Only one laboratory experiment had been
carried out for the SSFT\cite{otherexperiment} and none for the TFT,
until recently by Wang {\em et al.\/}\cite{Wangetal02}.

All deterministic theories were concerned with systems in phase space
consisting of many particles.  The experiment of Wang {\em et al.\/}
was carried out for a single Brownian particle which was dragged by
means of a uniformly moving harmonic potential generated by a laser
through a many particle molecular solvent.  This system differs in an
important aspect from the many particle systems in the phase space
considerations.  This is due to the fact that the (mesoscopic)
Brownian particle is much heavier than the surrounding fluid
particles, which makes it tractable in a different, much simpler
though approximate, way from the full dynamical systems treatment in
phase space mentioned above. In fact, the treatment generally applied
to such systems is via a Langevin equation for the stochastic motion
of the Brownian particle in a medium in real space, which is
characterized only by its friction with the particle and its
temperature. As a consequence the very complicated many particle
problem can be treated by a single particle Langevin equation, if it
is near equilibrium and on the level of Irreversible
Thermodynamics\cite{Sevickprivate02}.

However, for the investigation of the Fluctuation Theorems, an
additional difficulty is that the experiment considered here, uses a
time dependent force on the particle, since the Langevin equation
contains a laser-induced harmonic force on the particle, where the
position of the minimum of the harmonic potential changes in time.  As
a consequence, the treatments in Refs.~\cite{Kurchan98} and
\cite{LebowitzSpohn99} do not directly apply to this experiment.

Furthermore, the phase space treatments of dynamical systems have
always been such that (at least if the total energy of the system is
kept constant) the total phase space contraction can be directly
related to the total entropy production of the system. This has led to
the TFT\cite{Evansetal94,SearlesEvans00} and the
SSFT\cite{Evansetal93,GallavottiCohen95a,GallavottiCohen95b} for the
entropy production. To be sure, this connection between phase space
contraction and entropy production can only be made if the total work
done on the system is purely dissipative.  However, in the Wang {\em
et al.\/} experiment, this is not so.

To see this, it is useful to consider the total ($tot$) work
$W^{tot}_\tau$ done on the system during a time $\tau$:
\begin{equation}
    W^{tot}_\tau = \int_0^\tau\!dt\, \mathbf v^*_t\cdot
        \mathbf F(\mathbf x_t,\mathbf x^*_t),
\label{WS}
\end{equation}
where 
\begin{equation}
    \mathbf F(\mathbf x_t,\mathbf x^*_t) = -k(\mathbf x_t-\mathbf x^*_t).
\label{defF}
\end{equation}
is the harmonic force exerted on the particle, with $\mathbf x_t$ the
position of the particle and $\mathbf x^*_t$ the position of the
minimum of the harmonic potential, and $k$ the force constant of this
potential. Furthermore, in \Eq{WS} $\mathbf v^*_t=\dot{\mathbf
x}^*_t$.  At $t \leq 0$, in the Wang experiment, the center of the
harmonic potential is at rest at $\mathbf x^*_0=0$. At $t=0$, the
harmonic potential is set in motion relative to the fluid with a
constant velocity $\mathbf v^*$, so that $\mathbf x^*_t = \mathbf v^*
t$ for $t\geq0$\cite{footnote1}.

The crucial question is now what is the dissipative part of
$W^{tot}_\tau$ which is responsible for the heat or entropy produced
in the system as a result of the friction of the particle with the
surrounding fluid. The dissipative part should not include any purely
mechanical work. To see how $W^{tot}_\tau$ is related to the
dissipated work over a time $\tau$, we rewrite the total work done,
$W^{tot}_\tau$ in \Eq{WS}, as follows
\begin{eqnarray}
    W^{tot}_\tau &=& \int_0^\tau\!dt\, \mathbf v^*_t\cdot
        \mathbf F(\mathbf x_t,\mathbf x^*_t) 
\nonumber\\
&=& - \int_0^\tau\!dt\, (\mathbf v_t-\mathbf v^*_t)\cdot
        \mathbf F(\mathbf x_t,\mathbf x^*_t) 
+
\int_0^\tau\!dt\, \mathbf v_t\cdot
        \mathbf F(\mathbf x_t,\mathbf x^*_t)
\nonumber\\ 
&=& k \int_0^\tau\!dt\, (\dot{\mathbf x}_t -\dot{\mathbf x}^*_t )\cdot
            \left(\mathbf x_t-\mathbf x^*_t \right) 
+
\int_0^\tau\!dt\, \mathbf v_t\cdot
        \mathbf F(\mathbf x_t,\mathbf x^*_t)
\nonumber
\\
        &=& \Delta U
        + W^{Br}_\tau,
\label{WSWB}
\end{eqnarray}
defining $\Delta U\equiv \frac{k}{2}\left[|\Delta\mathbf x_\tau|^2
-|\Delta\mathbf x_0|^2\right]$ with $\Delta\mathbf x_t=\mathbf
x_t-\mathbf x^*_t$, and
\begin{equation}
	W^{Br}_\tau \equiv \int_0^\tau\!dt\, \mathbf v_t\cdot
        \mathbf F(\mathbf x_t,\mathbf x^*_t).
\end{equation}
Here, $W^{Br}_\tau$ is the work done on the Brownian ($Br$) particle
by the harmonic force. As (at least ideally) the Brownian particle has
no internal energy, all this work is converted into heat, which is the
source of the entropy production. Hence, $W^{Br}_\tau$ {\em is} the
dissipated work.  On the other hand, the term $\Delta U$ in \Eq{WSWB}
represents the purely mechanical (``center of mass'') work done on the
particle in the external harmonic potential.

Therefore, Wang {\em et al.\/}'s entropy production during time
$\tau$, denoted by $\Sigma_\tau$ in Ref.~\cite{Wangetal02}, is really
the total dimensionless work done on the system, and we will denoted
it by
\begin{equation}
	W_\tau = \beta W^{tot}_\tau = \beta
		\int_0^\tau\!dt\, 
		 \mathbf v^*_t\cdot
		\mathbf F(\mathbf x_t,\mathbf x^*_t) ,
\label{Sigmatau}
\end{equation}
where $\beta\equiv1/(k_BT)$, with $k_B$ Boltzmann's constant and $T$
the temperature of the surrounding fluid.  By following the position
of many, independent Brownian particles and using
\Eqs{defF}{Sigmatau}, Wang {\it et al.}  measured this dimensionless
work $W_\tau$ --- or what they called the entropy production
$\Sigma_\tau$ --- over time intervals $\tau$ and constructed from that
the probability distribution function $P(W_\tau)$, which they found
satisfies
\begin{equation}
	\frac{P(W_\tau)}{P(-W_\tau)} = e^{W_\tau}.
\label{TFT1}
\end{equation}
So they established experimentally the validity of a TFT for the total
work done on the system ($\beta W^{tot}_\tau$), rather than for the
entropy production of the system. Strictly speaking, Wang {\em et
al.\/}\ measured an integrated variant (an ITFT) of the TFT in
\Eq{TFT1}, explained in Sec.~\ref{IFT}.  A direct transformation of
the TFT \eq{TFT1} for $W_\tau$ to a TFT for the dimensionless entropy
production, which would be $\beta W^{Br}_\tau$, is not obvious, since
$\Delta U$ in \Eq{WSWB} fluctuates also.

We remark that Mazonka and Jarzynski\cite{MazonkaJarzynski99} studied
the same system as used in the experiment of Wang {\em et al.\/}
theoretically --- before the experiment --- and derived the TFT and
the SSFT for the total work done on the system, but not for the
entropy production\cite{footnote2}.

Unaware of Mazonka and Jarzynski's work, but in view of the experiment
of Wang {\em et al.\/}, we studied this experiment independently
\cite{VanZonCohen02a}.  For, Wang {\em et al.\/}'s experiment is
clearly important for practical purposes, since it involves a general
property of the work done on a system. We discuss the observability of
the work related TFT as well as SSFT for an arbitrary rather than a
uniform motion of the harmonic potential.  So, for the purpose of
treating the experiment and its generalizations, in this paper we too
will treat the TFT and SSFT for the dimensionless work, with a focus
on the feasibility to do a convincing SSFT experiment.  To the best of
our knowledge, no Fluctuation Theorem for {\em entropy production},
either an (I)TFT or an (I)SSFT, has been derived for a Wang-type
system, neither from a phase space perspective nor in real space (via
a Langevin equation).

The outline of the paper is as follows.  In Sec.~\ref{Theory} we
present our Langevin model and we develop the general theory for the
verification and the experimental observability of the work related
fluctuation theorems, and discuss a new relation between the
fluctuations in the transient and in the stationary state.  In
Sec.~\ref{Applications}, we specialize the general theory to the case
of a linear and a circular motion of the minimum of the harmonic
potential, investigating in detail the observability of the ITFT and
the ISSFT.  In Sec.~\ref{Discussion} we end with a discussion.

\section{Theory}
\label{Theory}

\subsection{Definition of the Model}

Like in the experiment of Wang, the model we consider has a spherical
Brownian particle in three dimensions with a radius $R$ and mass $m$
in a fluid with viscosity $\eta$ and temperature $T$ and the Brownian
particle is subject to an external harmonic potential with a time
dependent position $\mathbf x_t^*$ of its minimum.  For $t\leq0$ the
minimum of the harmonic potential is at the origin, $\mathbf x^*_t=0$,
whereas for $t>0$ it moves with a velocity $\mathbf v^*_t$, which can
be, in principle, an arbitrary function of time.  The equations of
motion for the particle are then of the Langevin-type:
\begin{subequations}
\begin{eqnarray}
    \dot{\mathbf x}_t &=& \mathbf v_t
\label{dotx}
\\
    m\dot{\mathbf v}_t &=& -\alpha\mathbf v_t
                -k(\mathbf x_t-\mathbf x^*_t)
                + \zeta_t,
\label{dotv}
\end{eqnarray}
\end{subequations}
where $\mathbf x_t$ and $\mathbf v_t$ are the position and velocity of
the Brownian particle, respectively. In this equation, the Brownian
particle feels three forces. The first force is the drag force
$-\alpha\mathbf v_t$, with according to Stokes' law
\begin{equation}
    \alpha=6\pi\eta R.
\label{Stokes}
\end{equation}
The second force is due to the harmonic potential [see \Eq{defF}].
The third and last force is a random force $\zeta_t$, which is taken
to be Gaussian and delta-correlated in time:
\begin{equation}
    \average{\zeta_t} = 0;
\quad
    \average{\zeta_t\zeta_s} = 2k_B T\alpha\delta(t-s),
\label{sf}
\end{equation}
The strength of the random force in \Eq{sf} is such that the
equilibrium distribution function $p_{eq}$ for $\mathbf x$ and
$\mathbf v$
\begin{equation}
   f_{eq}(\mathbf x,\mathbf v)
    = \left(\frac{\beta\sqrt{km}}{2\pi}\right)^{3}
            e^{
            -\beta\left(\frac12 m|\mathbf v|^2
            + \frac12 k|\mathbf x|^2\right)
            },
\label{equilibrium0}
\end{equation}
is stationary under the equations of motion
Eqs.(\ref{dotx}-b)\cite{UhlenbeckOrnstein}.

The system will only be considered in the strongly overdamped case
\begin{equation}
    mk\ll \alpha^2.
\label{overdamped}
\end{equation}
Effectively therefore, the mass can be seen as a small parameter and
will be set equal to zero\cite{UhlenbeckOrnstein}. From
\Eqs{dotx}{dotv}, we find then a simplified Langevin equation for the
position of the particle only
\begin{equation}
    \dot{\mathbf x}_t = -\tau_r^{-1} (\mathbf x_t-\mathbf x^*_t)
                + \alpha^{-1}\zeta_t,
\label{Langevin}
\end{equation}
with a relaxation time
\begin{eqnarray}
\tau_r &=& \frac{\alpha}{k}.
\label{taurdef}
\end{eqnarray}
When we only use $\mathbf x_t$, the equilibrium distribution in
\Eq{equilibrium0} reduces to
\begin{equation}
    p_{eq}(\mathbf x)
=\int\! d\mathbf v\, f_{eq}(\mathbf x,\mathbf v) 
    = (k\beta/2\pi)^{3/2}
            e^{-\beta\frac{k}{2}|\mathbf x|^2}.
\label{equilibrium}
\end{equation}

It is convenient to separate the average motion of the Brownian
particle (which results from the deterministic forces alone), from the
stochastic motion. The average motion is given by the solution
$\mathbf y^*_t$ of the deterministic part of the Langevin equation
\eq{Langevin}, i.e., by
\begin{equation}
    \dot{\mathbf y}^*_t = -\tau_r^{-1}(\mathbf y^*_t-\mathbf x^*_t).
\label{yt}
\end{equation}
with initial condition $\mathbf y^*_0=0$.  We can then look at the
deviations from this average motion by introducing the transformation
\begin{equation}
    \mathbf X_t = \mathbf x_t - \mathbf y^*_t.
\label{comoving}
\end{equation}
This turns the Langevin equation \eq{Langevin} into the simple form
\begin{equation}
    \dot{\mathbf X}_t = -\tau_r^{-1}\mathbf X_t + \alpha^{-1}
    \zeta_t.
\label{OU}
\end{equation}
$\mathbf y^*_t$ follows from the general solution of \Eq{yt}:
\begin{equation}
    \mathbf y^*_t = e^{- t/\tau_r} \mathbf y^*_0
    + \tau_r^{-1} \int_0^t\!dt'\, e^{-(t-t')/\tau_r}
                \mathbf x^*_{t'},
\end{equation}
so that with $\mathbf y^*_0=0$, and a partial integration, one obtains
\begin{equation}
    \mathbf y^*_t
    = \mathbf x^*_t - \int_0^t \!dt'\,e^{-(t-t')/\tau_r}\mathbf v^*_{t'}.
\label{ytex}
\end{equation}
The transformation \eq{comoving} with \Eq{ytex} can be interpreted as
going to a co-moving frame, but it is not co-moving with the minimum
of the harmonic potential, but with $\mathbf y^*_t$ which is what the
motion of a particle starting at $\mathbf x^*_0=0$ would be if there
would be no noise term in the Langevin equation \eq{Langevin}.

\Eq{OU} shows that in the co-moving frame, one has the standard
Ornstein-Uhlenbeck process\cite{VanKampen,UhlenbeckOrnstein}.  Its
solutions are well-known.  The Green's function of the
Ornstein-Uhlenbeck process, which gives the probability for the
particle to be at $\mathbf X_1$ at time $t_1$, given that it was at
$\mathbf X_0$ at time $t_0$, is Gaussian in both $\mathbf X_0$ and
$\mathbf X_1$. Its stationary solution is of the form $p_{eq}(\mathbf
X)$, with $p_{eq}$ given in \Eq{equilibrium}. Initially the particle
is distributed according to \Eq{equilibrium}, but because $\mathbf
X_0=\mathbf x_0$ ($\mathbf y^*_0=0$), one sees that the initial
distribution is already the stationary one, and in this special,
co-moving coordinate frame, the distribution of the Brownian particle
has an equilibrium distribution for all time:
\begin{equation}
P(\mathbf X,t)=(\beta k/2\pi)^{3/2}e^{-\beta\frac{k}{2}|\mathbf X|^2}.
\label{alltime}
\end{equation}

We end this section by writing $W_\tau$ in \Eq{Sigmatau} in terms of
$\mathbf X_t$,
\begin{equation}
    W_\tau = -k\beta \int_0^\tau \!dt\, \left[\mathbf
    v^*_t\cdot\mathbf X_t + \mathbf v^*_t\cdot(\mathbf y^*_t-\mathbf
    x^*_t)\right].  \label{SigmatauX}
\end{equation}

\subsection{Transient Fluctuation Theorem for the Total Work}

In \Eq{SigmatauX}, $W_\tau$ is a linear function of $\mathbf X_t$.
Combined with the Gaussian nature both of the Green's function of the
Ornstein-Uhlenbeck process [\Eq{OU}] and of the initial
distribution[\Eq{alltime}], this means that the distribution $P_T$ of
$W_\tau$ is Gaussian
\begin{equation}
    P_T(W_\tau) = \frac{
    e^{-\frac{[W_\tau-M_T(\tau)]^2}{2V_T(\tau)}}
    }
    {\sqrt{2\pi V_T(\tau)}},
\label{PT}
\end{equation}
where the subscript $T$ denotes that the transient case is
considered. The mean $M_T$ of $W_\tau$ is, from \Eq{SigmatauX},
\begin{equation}
    M_{T}(\tau) =
- k\beta \int_0^\tau \!dt\, \mathbf v^*_t\cdot(\mathbf y^*_t-\mathbf x^*_t),
\label{MTdef}
\end{equation}
since $\average{\mathbf X_t}=0$ [\Eq{alltime}].  Using the expression
for $\mathbf y^*_t$ in \Eq{ytex}, this can be also written as
\begin{equation}
    M_{T}(\tau) = k\beta \int_0^\tau\!dt_2'\int_0^{t_2'}\!dt'_1\,
    e^{-(t_2'-t_1')/\tau_r}
    \mathbf v^*_{t_2'}\cdot\mathbf v^*_{t_1'}
\label{MT}
\end{equation}
The variance $V_T$ of $W_\tau$ is only affected by the first term in
\Eq{SigmatauX}, so that
\begin{eqnarray}
    V_T(\tau) &=& \laverage{(W_\tau - \average{W_\tau})^2}
\nonumber\\
    &=& 2k^2\beta^2\int_0^\tau\!dt_2'\int_0^{t_2'}\!dt'_1\,
        \mathbf v^*_{t_2'}\cdot
        \average{\mathbf X_{t_2'}\mathbf X_{t_1'}}
        \cdot\mathbf v^*_{t_1'},
\label{VT}
\end{eqnarray}
where we used the symmetry of the time-correlation function
$\average{\mathbf X_{t_2'}\mathbf X_{t_1'}}$ under interchange of
$t'_1$ and $t'_2$.  To calculate this function, notice that $\mathbf
X_t$ has a stationary distribution so it can be written as
$\average{\mathbf X_{t'_2-t'_1}\mathbf X_0}$. Using the formal
solution of the Langevin equation in the co-moving frame [\Eq{OU}] for
$t>0$,
\begin{equation}
    \mathbf X_t = e^{-t/\tau_r} \mathbf X_0 + \alpha^{-1}\int_0^t\!dt'\,
    e^{-(t-t')/\tau_r}\zeta_{t'},
\end{equation}
one obtains with $\average{\zeta_{t'}}=0$, $\average{\mathbf X_0}=0$,
$\average{\zeta_{t'}\mathbf X_0}=0$ and $\average{\mathbf X_0\mathbf
X_0}=[k\beta]^{-1}\identity$,
\begin{equation}
\average{\mathbf X_t\mathbf X_0} = [\beta k]^{-1}e^{-t/\tau_r}\identity.
\end{equation}
The variance in \Eq{VT} then becomes
\begin{equation}
    V_T(\tau)
    = 2k\beta\int_0^\tau\!dt_2'\int_0^{t_2'}\!dt'_1\,
    e^{-(t_2'-t_1')/\tau_r} 
        \mathbf v^*_{t_2'}\cdot\mathbf v^*_{t_1'}
	.
\label{VT2}
\end{equation}
Comparing with the mean in \Eq{MT}, we see
\begin{equation}
    V_T(\tau) = 2 M_T(\tau).
\label{VTMT}
\end{equation}
This relation leads straightforwardly to the TFT: Given the
distribution function of $W_\tau$ in \Eq{PT}, one easily shows that
\begin{equation}
	\frac{P_T(W_\tau)}{P_T(-W_\tau)} = e^{
	\frac{2M_T(\tau)W_\tau}{V_T(\tau)}},
\end{equation}
which, by \Eq{VTMT}, becomes
\begin{equation}
	\frac{P_T(W_\tau)}{P_T(-W_\tau)} = e^{W_\tau},
\label{TFT}
\end{equation}
which is identical to the TFT in \Eq{TFT1}.

\subsection{Stationary State Fluctuation Theorem for the Total Work}

\label{SSFT}

To move on to the SSFT, it is necessary to clarify what the stationary
state means, since in the $\mathbf X$ coordinate system, the
distribution is stationary, which would suggest that the TFT is also
the SSFT. This is not the case. If one defines a stationary state as
that state in which (on average) the physical (macroscopic) parameters
do not change, then the time-independence of the distribution of
$\mathbf X$ is not enough because $\mathbf X$ involves, through its
definition \Eq{comoving}, a {\em time-dependent} transformation from
the laboratory frame, in which the physical parameters are measured,
so that they would still depend on time. Only when these parameters
have become stationary can one say that the system is stationary.

The SSFT was originally formulated for the average entropy production
fluctuations on trajectory segments of length $\tau$ along a single
trajectory in the stationary state. Here we consider the statistics of
the total work done on the system over time $\tau$, divided by $k_BT$,
\begin{equation}
    W_\tau = \beta\int_{t_i}^{t_i+\tau} \!dt\,\mathbf v^*_t\cdot 
	\mathbf F(\mathbf x_t,\mathbf x_t^*),
\label{Sigmatauu}
\end{equation}
for a sequence of initial times $t_i$ of segments, all of length
$\tau$, along a single stationary state trajectory
($i=1,2,3,\ldots$). To get the distribution of $W_\tau$ of the
segments along a trajectory, we use the following reasoning. According
to the Eqs. (\ref{defF}) (for the force) and \eq{comoving} (the
definition of $\mathbf X$), the expression in \Eq{Sigmatauu} is linear
in $\mathbf X_t$ (just as in the transient case) and [as $\mathbf X_t$
obeys the Langevin equation \eq{OU}] we still have a Gaussian Green's
function and a Gaussian stationary state, so that the distribution of
$W_\tau$ {\em for each $t_i$} is again Gaussian:
\begin{equation}
    P_{t_i}(W_\tau) =
    \frac{e^{-\frac{
          [W_\tau-M_{t_i}(\tau)]^2
          }
         {2V_{t_i}(\tau)}
           }}
    {\sqrt{2\pi V_{t_i}(\tau)}
},
\label{Pu}
\end{equation}
with the mean and the variance given by, respectively:
\begin{eqnarray}
M_{t_i}(\tau) &=&
- k\beta \int_{t_i}^{t_i+\tau} \!dt\,\mathbf v^*_t\cdot(\mathbf y^*_t-\mathbf x^*_t),
\label{Mu}
\\
    V_{t_i}(\tau)
    &=& 2k\beta\int_{t_i}^{t_i+\tau} \!dt_2'\int_{t_i}^{t_2'} \!dt'_1\,
        \mathbf v^*_{t_2'}\cdot\mathbf v^*_{t_1'}
        e^{-(t_2'-t_1')/\tau_r}.
\label{Vu}
\end{eqnarray}
We assume that for sufficiently large $t_i$, $M_{t_i}$ and $V_{t_i}$
will reach steady state values (see Sec.~\ref{Applications} for
examples), and become independent of $i$. If in addition, the
correlation between different segments ($[t_i,t_i+\tau]$ and
$[t_j,t_j+\tau]$, say) decays sufficiently fast (when $|t_i-t_j|$ gets
larger), then the distribution of $W_\tau$ along a trajectory in the
stationary state is given by:
\begin{equation}
    P_S(W_\tau) =
    \frac{
    e^{-\frac{\left[W_\tau-M_S(\tau)\right]^2}
                {2V_S(\tau)}
        }
	}
    {\sqrt{2\pi V_S(\tau)}}.
\label{PS}
\end{equation}
Here the subscript $S$ denotes that this distribution refers to the
distribution of $W_\tau$ over segments along the stationary state
trajectory. The mean $M_S$ is, from \Eq{Mu} and using \Eq{ytex}, given
by: \newcommand{\tw}{t}
\begin{equation}
	M_S(\tau) =\lim_{\tw\rightarrow\infty}
	k\beta \int_{\tw}^{\tw+\tau} \!dt_2'
	\int_0^{t_2'}\!dt'_1\,
	e^{-(t_2'-t_1')/\tau_r}	
	\mathbf v^*_{t_2'}\cdot\mathbf v^*_{t_1'},
\label{MS}
\end{equation}
while the variance $V_S$ is, from \Eq{Vu}, given by
\begin{equation}
    V_S(\tau)
    = \lim_{\tw\rightarrow\infty}
    2k\beta\int_{\tw}^{\tw+\tau}\!dt_2'\int_{\tw}^{t_2'}\!dt'_1\,
        e^{-(t_2'-t_1')/\tau_r}
        \mathbf v^*_{t_2'}\cdot\mathbf v^*_{t_1'}
       .
\label{VS}
\end{equation}
Note that in the inner most integral in the expression for the mean in
\Eq{MS}, the lower bound extends to time zero, whereas in the
expression for the variance in \Eq{VS}, it extends to $\tw$. This is
the origin of the fact that $V_S$ and $2M_S$ are not identical [while
$V_T=2M_T$, \Eq{VTMT}]. The deviation can be characterized by
\begin{eqnarray}
    \varepsilon(\tau) &\equiv&
    \frac{2M_S(\tau)-V_S(\tau)}{2M_S(\tau)}
\label{epsilondef}
\end{eqnarray}
Using this definition and \Eq{PS}, one sees that
\begin{eqnarray}
\frac{P_S(W_\tau)}{P_S(-W_\tau)}
&=& \exp\left\{\frac{W_\tau}{1-\varepsilon(\tau)}\right\}.
\label{almostSFT}
\end{eqnarray}
This means that provided that
\begin{equation}
\varepsilon(\tau)
\rightarrow 0 \mbox{ as } \tau\rightarrow\infty,
\label{E}
\end{equation}
we have 
\begin{equation}
    V_S(\tau) \rightarrow 2 M_S(\tau)  \mbox{ as } \tau\rightarrow\infty,
\label{VSMS}
\end{equation}
and the SSFT holds
\begin{equation}
    \frac{P_S(W_\tau)}{P_S(-W_\tau)}
    \rightarrow e^{W_\tau}  \mbox{ as } \tau\rightarrow\infty.
\label{eSSFT}
\end{equation}

Of course, for any given $\mathbf v^*_t$, \Eq{E} can be tested, but
how general can we expect it to be satisfied?  We write thereto
\Eq{epsilondef} with \Eqs{MS}{VS} as:
\begin{eqnarray}
\varepsilon(\tau)
&=& \frac{\displaystyle \lim_{\tw\rightarrow\infty}k\beta \int_{\tw}^{\tw+\tau}\!dt_2'
\int_0^{\tw}\!dt'_1\,
 \mathbf v^*_{t'_2}\cdot\mathbf v^*_{t_1'}
e^{-(t'_2-t'_1)/\tau_r}}{M_S(\tau)}
\nonumber\\
&=&
\frac{\displaystyle\lim_{\tw\rightarrow\infty}
	k\beta (\mathbf x^*_{\tw}-\mathbf y^*_{\tw})
	\int_{0}^{\tau} \!dt_2'\, e^{-t_2'/\tau_r} \mathbf v^*_{\tw+t_2'}
	}{M_S(\tau)},
\label{epsilon}
\end{eqnarray}
where \Eq{ytex} has been used.  Here, the denominator is the total
work done of the system in the stationary state in time $\tau$. If we
are not in equilibrium, this is positive and grows with $\tau$. In the
numerator, the exponential in the integral will make the integral
bounded for large $\tau$, provided that $\mathbf v^*_t$ does not grow
exponentially in time with an exponent bigger than $\tau_r^{-1}$.
Then $\varepsilon$ will become zero $\propto 1/\tau$ as $\tau$
approaches infinity, and the SSFT in \Eq{eSSFT} holds.

\subsection{Integrated Fluctuation Theorems}
\label{IFT}

In experiments such as done by Wang {\em et al.}\cite{Wangetal02}, it
is easier to check an integrated fluctuation
theorem\cite{Aytonetal01}, because it is easier to obtain then good
statistics for the required quantities. The Integrated Transient
Fluctuation Theorem (ITFT) reads
\begin{subequations}
\begin{equation}
    \frac{P_T(W_\tau<0)}{P_T(W_\tau>0)}
=    \laverage{e^{-W_\tau}}_{T}^{^{_{+}}},
\label{ITFT}
\end{equation}
where the left hand side is the quotient of the probabilities to see a
negative resp. a positive total work $W_\tau$ after a time $\tau$:
\begin{eqnarray}
    P_T(W_\tau<0)
    &\equiv&
    \int_{-\infty}^0\!dW_\tau\, P_T(W_\tau) \\
    P_T(W_\tau>0)&=&1-P_T(W_\tau<0)
\end{eqnarray}
and the right hand side of \Eq{ITFT} is the average of
$\exp(-W_\tau)$ over positive $W_\tau$, i.e.,
\begin{eqnarray}
    \laverage{e^{-W_\tau}}_{T}^{^{_{+}}} &\equiv& \frac
    {\int_0^\infty \!dW_\tau\,P_T(W_\tau)e^{-W_\tau}}
    {\int_0^\infty \!dW_\tau\,P_T(W_\tau)}.
\end{eqnarray}
\end{subequations}

The ITFT of \Eq{ITFT} can be derived from the TFT in \Eq{TFT} by first
rewriting $P_T(W_\tau<0) =\int_{-\infty}^{0}\! dW_\tau\, P_T(W_\tau)$
as
\begin{eqnarray}
    \int_{-\infty}^{0}\! dW_\tau\, P_T(W_\tau)
&=& \int_{-\infty}^{0}\! dW_\tau\, P_T(-W_\tau) e^{W_\tau}
\nonumber\\
&=& \int_0^{\infty} \! dW_\tau\,P_T(W_\tau) e^{-W_\tau} ,
\end{eqnarray}
and then dividing by $P_T(W_\tau>0)$.

An Integrated Stationary State Fluctuation Theorem (ISSFT) can also be
derived, but it is a little more subtle. Thereto, one has to consider
whether
\begin{subequations}
\begin{equation}
    \frac{P_S(W_\tau<0)}{P_S(W_\tau>0)}
    \stackrel{\tau\rightarrow\infty}{=}
     \laverage{e^{-W_\tau}}_{S}^{^{_{+}}}
\label{ISFT}
\end{equation}
holds, where
\begin{eqnarray}
    P_S(W_\tau<0) &\equiv& \int_{-\infty}^{0}\! dW_\tau\, P_S(W_\tau)
    \\ P_S(W_\tau>0) &\equiv& 1-P_S(W_\tau<0),
\end{eqnarray}
and
\begin{equation}
    \laverage{e^{-W_\tau}}_{S}^{^{_{+}}}
    \equiv
     \frac
    {\int_0^\infty\! dW_\tau\, P_S(W_\tau)e^{-W_\tau}}
    {\int_0^\infty\! dW_\tau\, P_S(W_\tau) }.
\label{Sstat}
\end{equation}
\end{subequations}
To start the derivation of the ISSFT of \Eq{ISFT}, the numerator of
\Eq{Sstat} is rewritten, using \Eq{almostSFT}, as
\begin{eqnarray}
&&  \int_0^\infty\! dW_\tau\, P_S(W_\tau)e^{-W_\tau} 
\nonumber\\&& = \int_0^\infty\! dW_\tau\, P_S(-W_\tau)\exp\left\{
    \frac{\varepsilon(\tau)W_\tau}{1-\varepsilon(\tau)}
    \right\}
\nonumber\\&& = \int_{-\infty}^0\! dW_\tau\, P_S(W_\tau)\exp\left\{-
    \frac{\varepsilon(\tau)W_\tau}{1-\varepsilon(\tau)}
    \right\} 
\nonumber\\
&&=
\int_{-\infty}^0\frac{
    \EXP{-\frac{\left[W_\tau-M_S(\tau)\right]^2}
                {2V_S(\tau)}
    -
    \frac{\varepsilon(\tau)}{1-\varepsilon(\tau)}
    W_\tau
        }
     }
     {\sqrt{2\pi V_S(\tau)}},
\label{almostthere}
\end{eqnarray}
where \Eq{PS} was used.  We saw that the SSFT holds if
$\varepsilon\rightarrow 0$ for large $\tau$.  Consider the exponent in
\Eq{almostthere}. Writing out the square, this has a term linear in
$W_\tau$ of the form:
\begin{equation}
\left[\frac{M_S(\tau)}{V_S(\tau)}-
    \frac{\varepsilon(\tau)}{1-\varepsilon(\tau)}\right]W_\tau
= \frac{M_S(\tau)}{V_S(\tau)}[1-2\varepsilon(\tau)]W_\tau.
\end{equation}
As $\tau\rightarrow\infty$, we can neglect $\varepsilon$ compared to
1. Since this is the only place where $\varepsilon$ occurs, we can set
$\varepsilon$ equal to zero on the right hand side of
\Eq{almostthere}, which then becomes $P(W_\tau<0)$. Dividing by
$P(W_\tau>0)$ on both sides in \Eq{almostthere} now yields the ISSFT
in \Eq{ISFT}.

For the purpose of the investigation of the observability of the
fluctuation theorems, discussed in Sec.~\ref{Applications}, we end
this section by giving the explicit forms of the left and right hand
sides of the integrated fluctuation theorems \Eq{ITFT} and \Eq{ISFT}.
Defining
\begin{subequations}
\begin{eqnarray}
   L_T
   (\tau) \equiv \frac{P_T(W_\tau <0)}{P_T(W_\tau >0)}
\label{RT}
;&&
   R_T
   (\tau) \equiv \laverage{e^{-W_\tau}}_{T}^{^{_{+}}};
\label{ST}
\\
   L_S
  (\tau) \equiv \frac{P_S(W_\tau <0)}{P_S(W_\tau >0)}
\label{RS}
;&&
   R_S
   (\tau) \equiv \laverage{e^{-W_\tau}}_{S}^{^{_{+}}},
\label{SS}
\end{eqnarray}
\end{subequations}
the TFT states that $L_T=R_T$, and the SSFT that $L_S=R_S$ (the latter
for large $\tau$ only). Using Eqs.(\ref{PT}), \eq{PS} and \eq{ITFT},
we get the following explicit expressions:
\begin{subequations}
\begin{eqnarray}
   L_T
   (\tau) &=& R_T(\tau)
\label{STex0}
\\
   R_T
   (\tau) &=&
\label{RTex}
\frac{1-\erf\left(\frac{M_T(\tau)}{\sqrt{2V_T(\tau)}}\right)}
     {1+\erf\left(\frac{M_T(\tau)}{\sqrt{2V_T(\tau)}}\right)}
\\
   L_S
   (\tau) &=&
\frac{1-\erf\left(\frac{M_S(\tau)}{\sqrt{2V_S(\tau)}}\right)}
     {1+\erf\left(\frac{M_S(\tau)}{\sqrt{2V_S(\tau)}}\right)}
\label{RSex}
\\
   R_S
   (\tau) &=&
e^{\frac{V_S(\tau)}{2}-M_S(\tau)}
\frac{1-\erf\left(\frac{V_S(\tau)-M_S(\tau)}
    {\sqrt{2V_S(\tau)}}\right)}
     {1+\erf\left(\frac{M_S(\tau)}{\sqrt{2V_S(\tau)}}\right)}
\label{SSex}
\end{eqnarray}
\end{subequations}
We can simplify the expressions for $L_T$ and $R_T$ using the relation
between $M_T$ and $V_T$ in \Eq{VTMT}:
\begin{equation}
L_T
(\tau) = \frac{1-\erf\left(\frac12\sqrt{M_T(\tau)}\right)}
     {1+\erf\left(\frac12\sqrt{M_T(\tau)}\right)}.
\label{RTex2}
\end{equation}
In order to demonstrate the difference between $L_S$ and $R_S$, we
rewrite \Eq{SSex}, using \Eq{epsilondef}, in terms of $\varepsilon$ as
\begin{equation}
R_S
(\tau) = 
e^{-\varepsilon(\tau)M_S(\tau)}
\frac{1-\erf\left(\frac{[1-2\varepsilon(\tau)]M_S(\tau)}
    {\sqrt{2V_S(\tau)}}\right)}
     {1+\erf\left(\frac{M_S(\tau)}{\sqrt{2V_S(\tau)}}\right)}
\label{SSex2},
\end{equation}
which shows that only for $\tau\rightarrow\infty$, $L_S=R_S$, i.e.,
that the work related ISSFT holds.

\subsection{Transient Fluctuations versus Stationary Fluctuations}
\label{TSrelation}

An interesting relation can be derived for the ratio of the
probability of a negative total work and that of a positive one, for
the transient case ($L_T$) and the stationary case ($L_S$). Using
Eqs.(\ref{STex0}-c) and the asymptotic expansion of the error
function,
\begin{equation}
    \erf(x) = 1 - \frac{e^{-x^2}}{\sqrt{\pi}}
        \left[x^{-1}+{\cal O}(x^{-2})\right],
\label{erfasymptotic}
\end{equation}
one obtains 
\begin{equation}
    \frac{L_T}{L_S}
    \rightarrow
    \sqrt{\frac{V_T}{V_S}}\frac{M_S}{M_T}
	e^{-\frac{M_T^2(\tau)}{2V_T(\tau)}+\frac{M_S^2(\tau)}{2V_S(\tau)}}.
\label{startrel}
\end{equation}
Here the ${\cal O}(x^{-2})$ in \Eq{erfasymptotic} could be neglected.
This, because when $\tau\rightarrow\infty$, $M_T$ and $M_S$
[\Eqs{MT}{MS} respectively] will both grow linearly in time, $M_T\sim
M_S\sim{\cal O}(\tau)$ and similarly $V_T\sim V_S\sim{\cal O(\tau)}$
[by \Eqs{VTMT}{VSMS}], so that the arguments of the error functions in
Eqs.\eq{RTex} and \eq{RSex} become large ($\sim\sqrt\tau$) for large
$\tau$.  In fact, $M_S$ and $M_T$ will grow with the same coefficient,
as the rate of work done on the system will become stationary, but
they will in general have a bounded difference, i.e., $M_S-M_T\sim
{\cal O}(1)$, as will $V_S$ and $V_T$, i.e., $V_S-V_T\sim{\cal
O}(1)$. One therefore has, using \Eq{VTMT} and \Eq{VSMS}, an
asymptotic behavior
\begin{eqnarray}
	M_T(\tau) \rightarrow w \tau + a_1  ;&\quad&
	V_T(\tau) \rightarrow 2w \tau + 2 a_1 \nonumber\\
	M_S(\tau) \rightarrow w\tau + a_2 ;&\quad&
	V_S(\tau) \rightarrow 2w\tau + a_3 
\label{useme}
\end{eqnarray}
for $\tau\rightarrow\infty$, where the $a_i$ are independent of
$\tau$. $w$ is in fact the asymptotic rate at which work is supplied
to the system, i.e., the consumed power.  By expanding in terms of
$1/\tau$, we obtain for the exponent in \Eq{startrel} :
$-[M_T^2(\tau)/2V_T(\tau)] + [M_S^2(\tau)/[2V_S(\tau)]
=-\frac14a_1+\frac12a_2-\frac18a_3$, i.e., it approaches a non-zero
constant. The prefactor in \Eq{startrel} can similarly be shown to go
to one as $\tau\rightarrow\infty$, so that we have for large $\tau$,
\begin{eqnarray}
    \frac{L_T(\tau)}{L_S(\tau)}
&    \stackrel{\tau\rightarrow\infty}{=}&
    e^{ -\frac14a_1+\frac12a_2-\frac18a_3} 
\nonumber\\&    = &
    e^{\frac{M_S(\tau)-M_T(\tau)}{2}-\frac{V_S(\tau)-V_T(\tau)}{8}},
\label{tvs}
\end{eqnarray} 
where \Eq{useme} has been used to re-express this ratio in terms of
the means and variances of the transient, resp. stationary state.
Because the right hand side of \Eq{tvs} is not equal to $1$ in
general, this shows that the ISSFT is not just the limit of the ITFT
for large $\tau$.

\section{Applications}
\label{Applications}

In this section, two kinds of motion are considered for the harmonic
potential.  In both cases parameters will be varied to see under what
conditions an experiment would be able to demonstrate the integrated
work fluctuation theorems (both transient and stationary) most
convincingly.  The motion that will be considered first, corresponds
to the situation in the experiment of Wang {\em et al.}, i.e., it is a
uniform linear motion. The other is a circular motion and might be
implemented in a future experiment.  Both approach a stationary state.

\subsection{Linear Motion of the Harmonic Potential}
\label{sublinear}

The particular case considered here is a linearly moving harmonic
potential, i.e., $\mathbf x_t^*=v_{opt}t\mathbf{\hat x}$ for $t\geq0$.
The quantities with which to test the work fluctuation theorems, are
given in Eqs.~(\ref{STex0}--d). The only unknowns are the means and
variances of the transient and stationary distributions and these are
given by Eqs.~\eq{MT}, \eq{VTMT}, \eq{MS} and \eq{VS}.  If we insert
$\mathbf v^*_t$, which is a constant $v_{opt}\mathbf{\hat x}$ here,
into these equations, we obtain straightforwardly
\begin{subequations}
\begin{eqnarray}
    M_T(\tau) &=& w
        \left\{\tau - \tau_r[1-e^{-\tau/\tau_r}]
        \right\}
\label{MTlin}
\\
    V_T(\tau) &=& 2M_T(\tau)
\\
    M_S(\tau) &=& w\tau
\label{MSlin}
\\
    V_S(\tau) &=& V_T(\tau)
        =2 w
        \left\{\tau - \tau_r[1-e^{-\tau/\tau_r}]
        \right\},
\label{VSlin}
\end{eqnarray}
where
\begin{equation}
    w=\alpha\beta v_{opt}^2,
\label{sigma}
\end{equation}
\end{subequations}
which can be interpreted according to \Eq{MSlin} as the rate of work
delivered to the system per unit time.  The equality between $V_S$ and
$V_T$ in \eq{VSlin} follows because the velocity of the center of the
harmonic potential $\mathbf v^*_t$ is constant, so that the integrands
in \Eqs{VT2}{VS} only depend on the difference $t'_2-t'_1$, and the
shift over $\tw$ in the definition of $V_S$ is irrelevant.  For this
case, Eqs.~(\ref{MTlin}-d) were already derived by Mazonka and
Jarzynski in Ref.~\cite{MazonkaJarzynski99}.

By the theory presented in the Sec.\ref{Theory}, the TFT holds for any
motion of the harmonic potential, hence also in this case $L_T=R_T$.
The SSFT holds if $\varepsilon$ [\Eq{epsilondef}] vanishes as
$\tau\rightarrow\infty$, and this is so here, since
\begin{equation}
    \varepsilon(\tau) = \frac{\tau_r[1-e^{-\tau/\tau_r}]}{\tau}.
\label{lineps}
\end{equation}

\begin{figure}
\centerline{\includegraphics[width=\figwl\textwidth,angle=-90]{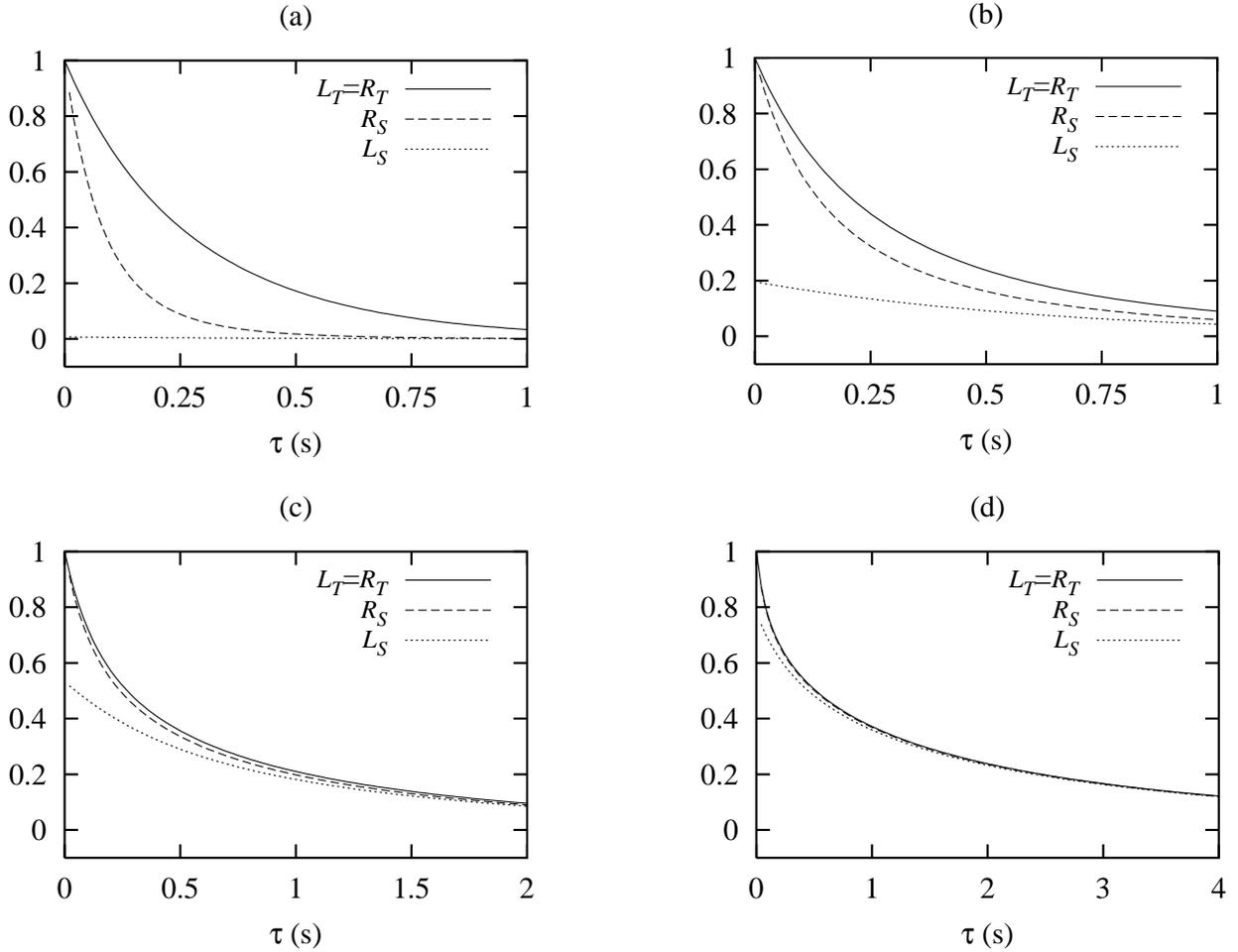}}
\caption{Integrated Fluctuation Theorems for the work for the linearly
moving harmonic potential: $L_T=R_T$ [cf. \Eq{STex0}], $R_S$
and $L_S$ versus $\tau$ (varying ranges) with (a) $\tau_r = 0.5\,
\rm s$ and $w = 12\, \rm s^{-1}$, (b) $\tau_r = 0.2\, \rm s$ and
$w = 4.8\, \rm s^{-1}$, (c) $\tau_r = 0.08\, \rm s$ and $w =
1.92\, \rm s^{-1}$, and (d) $\tau_r = 0.032\, \rm s$ and $w =
0.768\, \rm s^{-1}$ (in the last case the curves of $L_T=R_T$
and $R_S$ are indistinguishable).  }
\label{fig1}
\end{figure}

We now discuss the observability of the work related ITFT and the
ISSFT for this model.  There are only two relevant parameters for the
fluctuation theorems here, the relaxation time $\tau_r$ [\Eq{taurdef}]
and the rate of dimensionless work done $w$ [\Eq{sigma}].  To obtain
realistic values for these parameters, orders of magnitude of various
quantities can be taken from Ref.~\cite{Wangetal02}.  With the radius
$R$ of the order of three microns, and the viscosity of water $\eta$
of the order of $10^{-3}\,\rm kg\,m^{-1}s^{-1}$, according to
\Eq{Stokes}, $\alpha$ is of the order of $5\times 10^{-8}\rm
kg/s$. With $k$ of the order of $10^{-7}\,\rm kg\,s^{-2}$, $\tau_r$
becomes of the order of $0.5\,\rm s$ [\Eq{taurdef}].  Furthermore,
taking the temperature to be 300 K, gives $\beta$ of the order of
$2.4\times10^{20}\,\rm kg\, m^2\,s^{-2}$, and with $v_{opt}$ of the
order of $1 \,{\mu m/s}$, we find from \Eq{sigma} that $w$ is of the
order of $12\,\rm s^{-1}$.

For the case that $w=12\,\rm s^{-1}$ and $\tau_r=0.5\,\rm s$, the
expressions in Eqs.~(\ref{STex0}-d) using Eqs.~(\ref{MTlin}-d), are
plotted together in Fig.~\ref{fig1}a.  It is striking that even though
we know that the ISSFT holds for sufficiently large $\tau$, this is
not at all observed in the figure: the curves of $L_S$ and $R_S$ are
completely different; in fact, the curve for $L_S$ is
indistinguishable from the $\tau$ axis. Furthermore, both of these
curves are different from $L_T$, which, given the result in \Eq{tvs}
of Sec.~\ref{TSrelation}, is less of a surprise. Clearly, the range of
$\tau$ for which the ISSFT is valid lies beyond the point where both
$L_S$ and $R_S$ have relaxed to zero in Fig~\ref{fig1}a. This means
that {\em for the parameters typical of the Wang experiment, the ISSFT
cannot be observed}, in contrast to the ITFT, as the curve of $L_T$
can be seen clearly, and $L_T$ equals $R_T$, so that the ITFT could be
observed.

\begin{figure}[th]
\centerline{\includegraphics[angle=-90,width=\figwl\textwidth]{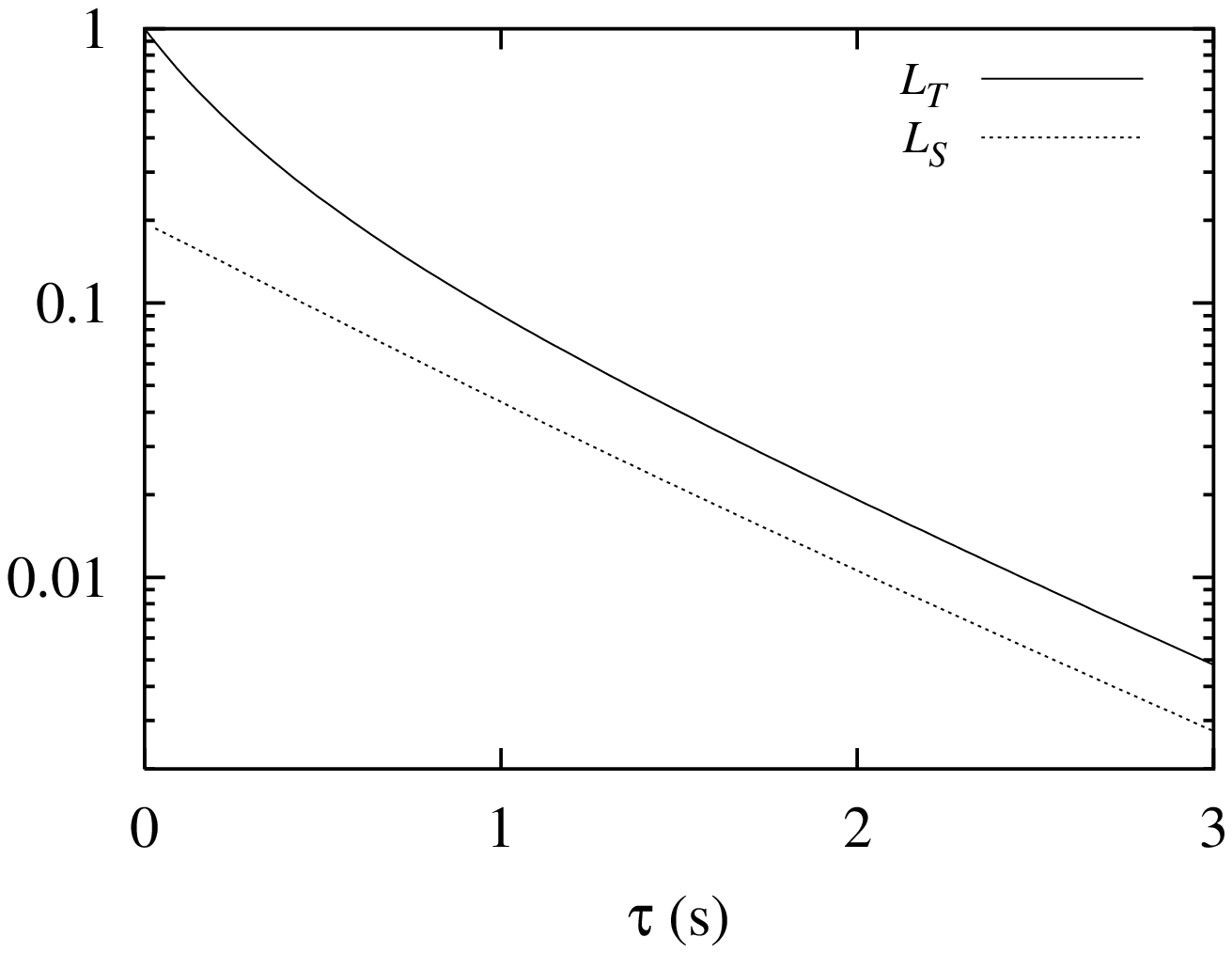}}
\caption{Illustration of the relation between transient and stationary
work fluctuations [\Eq{tvs}], for the linearly moving harmonic
potential.  $L_T$ and $L_S$ are plotted logarithmically as a
function of $\tau$, with $\tau_r = 0.2\, \rm s$ and $w = 4.8\,
\rm s^{-1}$ [cf. Fig.~\ref{fig1}b]. The constant ratio in \Eq{tvs}
becomes a constant distance between the curves in this logarithmic
plot.}
\label{fig2}
\end{figure}

The reason that there is such a big difference in the signal for the
transient and the stationary case, i.e., that the negative work
fluctuations are more suppressed in $L_S$ than in $L_T$, is the
following.  Since $M_T(0)=0$ [\Eq{MTdef}] in the argument of the error
functions in \Eq{RTex2}, $L_T(\tau=0)=1$.  As the function $L_T$
decays with increasing $\tau$, the question of whether it can be
observed depends on whether it does not decay too quickly.  On the
other hand, the argument of the error function in \Eq{RSex},
$M_S/\sqrt{2V_S}$ does not have a limit of zero for
$\tau\rightarrow0$. In fact, using \Eqs{MSlin}{VSlin} and expanding in
$\tau$, one finds
\begin{equation}
    \lim_{\tau\rightarrow 0} \frac{M_S(\tau)}{\sqrt{2V_S(\tau)}}
    = \sqrt{{w\tau_r}/{2}},
\label{limo}
\end{equation}
which can  be  large. So with \Eq{RSex}, one has for $\tau\rightarrow0$:
\begin{equation}
   L_S(0) 
  = \frac{1-\erf(\sqrt{{w\tau_r}/{2}})}
        {1+\erf(\sqrt{{w\tau_r}/{2}})}.
\label{Rzero}
\end{equation}
For the case plotted in Fig.~\ref{fig1}(a),
$\sqrt{w\tau_r/2}=\sqrt{3}$, so that $L_S(0)=7\times 10^{-3}$.  No
wonder we cannot see it in Fig.~\ref{fig1}(a). $L_S$ is exponentially
suppressed for large values of $w\tau_r$, which is the average work
done in the stationary state during a relaxation time $\tau_r$.  As
$W_\tau=0$ for $\tau=0$ for all trajectories, it follows from
definition \eq{SS} that $R_S(0)=1$.  Comparing this value with the
value of $L_S(0)$ [\Eq{Rzero}], we see that there is no chance to
observe the SSFT if the work done during $\tau_r$ is too large.

Thus in order to observe the SSFT we have to reduce the work done in
the time $\tau_r$. One direct way to do this is to change the
particle's radius. From Eqs.~\eq{Stokes} and \eq{taurdef}, we see that
$\tau_r\propto R$, whereas from \Eq{sigma}, $w\propto R$ as well, so
$w\tau_r\propto R^2$. Another way would be to reduce the velocity of
the harmonic potential, which does not affect $\tau_r$, but changes
$w\propto v_{opt}^2$. Choosing to work with the radius as the control
parameter (partly because smaller particles than used in the Wang
experiment are commercially available), we plotted in
Fig.~\ref{fig1}b--d, what happens when we make the particle 2.5 times
smaller consecutively, thus reducing the work done per relaxation time
each time by a factor $6.25$.  In first instance [Figs.~\ref{fig1}b
and \ref{fig1}c] the curve of $L_S$ gets closer to the curve $R_T$,
and $L_S$ starts to get visible. But only in the last graph,
Fig.~\ref{fig1}d, where the particle's diameter is about $15$ times
smaller than in Fig.~\ref{fig1}a (which would mean $R\approx 200\,\rm
nm$ in the Wang experiment), do we clearly see that $L_S$ approaches
$R_S$.  In this case, $w\tau_r\approx 0.02$, confirming that the work
done in $\tau_r$ needs to be small to see a convincing signal of the
ISSFT.

Finally, we look at \Eq{tvs}, which says that in the long $\tau$
limit, the ratio of $L_T$ and $L_S$ becomes a constant. Using
Eqs.~(\ref{MTlin}-d) in this case, \Eq{tvs} reads
\begin{equation}
    \frac{L_T(\tau)}{L_S(\tau)}
=   e^{\frac{w\tau_r}{2}}.
\label{svtlin}
\end{equation}
Taking $\tau_r=0.2\,\rm s$ and $w=4.8\,\rm s^{-1}$, we plotted $L_T$
and $L_S$ logarithmically to illustrate \Eq{svtlin}.  \Eq{svtlin}
shows once more that negative fluctuations of the work are more
suppressed in the stationary state than in the transient period.

\subsection{Circular Motion of the Harmonic Potential}
\label{subcircular}

In the case  of a circular motion of the harmonic potential, we write
\begin{equation}
    \mathbf x^*_t=r\{\sin(\Omega t)\mathbf{\hat x}+[1-\cos(\Omega
t)]\mathbf{\hat y}\},
\end{equation}
for $t\geq0$.  We use the same procedure as in the previous case,
i.e., we determine $\mathbf v^*_t$
\begin{equation}
	\mathbf v^*_t= r\Omega\{\cos(\Omega t)\mathbf{\hat
	x}+\sin(\Omega t) \mathbf{\hat y}\},
\end{equation}
for $t\geq0$, insert this into Eqs.~\eq{MT}, \eq{MS} and \eq{VS}, and
obtain straightforwardly
\begin{subequations}
\begin{eqnarray}
    M_T(\tau) &=&
        w
        \Bigg\{
        \tau
        - \tau_r\frac{2\Omega\tau_r\sin(\Omega\tau)e^{-\tau/\tau_r}}
        {1+\Omega^2\tau_r^2}
\label{circleMT}
\nonumber\\&&\:
        -\tau_r
        \frac{[1-\Omega^2\tau_r^2][1-\cos(\Omega\tau)e^{-\tau/\tau_r}]}
            {1+\Omega^2\tau_r^2}
        \Bigg\}
\\
    V_T(\tau) &=& 2M_T(\tau)
\\ M_S(\tau) &=& w\tau
\label{circleMS}
\\ V_S(\tau)
    &=& V_T(\tau) = 2 M_T(\tau),
\label{circleVS}
\end{eqnarray}
where now
\begin{equation}
w = \frac{\alpha\beta r^2\Omega^2}
                   {1+\Omega^2\tau_r^2}.
\label{sigmacircle}
\end{equation}
\end{subequations}
The expression for $\varepsilon$ follows using its definition
\Eq{epsilondef},
\begin{eqnarray}
    \varepsilon(\tau) &=& \frac{\tau_r}{(1+\Omega^2\tau_r^2)\tau}
        \Bigg\{
        2\Omega\tau_r\sin(\Omega\tau)e^{-\tau/\tau_r}
\nonumber\\&&
        +
        [1-\Omega^2\tau_r^2][1-\cos(\Omega\tau)e^{-\tau/\tau_r}]
    \Bigg\}
\label{epscircle}
\end{eqnarray}
which again vanishes like $1/\tau$ when $\tau\rightarrow\infty$, so
that the ISSFT holds, i.e., $L_S=R_S$ for large $\tau$.

Note that the Eqs.~(\ref{circleMT}-e) and (\ref{epscircle}) reproduce
the Eqs. (\ref{MTlin}-e) and (\ref{lineps}) in the limit for
$\Omega\rightarrow 0$, keeping $v_{opt}=r\Omega$ constant.  But these
equations are not just an extension of the linear case.  Under the
{\em resonance} condition $\Omega\tau_r=1$, $\varepsilon$ in
\Eq{epscircle} becomes
\begin{equation}
    \varepsilon(\tau) = \frac{\tau_r
    \sin({\tau}/{\tau_r})}{\tau} e^{-\tau/\tau_r},
\end{equation}
i.e., it decays exponentially, rather than $\propto 1/\tau$ (which it
does for all other choices of $\Omega$).  In addition, in the
resonance case $\varepsilon$ is zero at times $n\pi\tau_r$ for
$n=1,2,3,\ldots$, and exponentially small ($\sim e^{-n\pi}$) in
between; consequently, at those times, $L_S$ and $R_S$ are equal
[\Eqs{RSex}{SSex2}], whereas they are exponentially close in
between. This means that the ISSFT holds and can be visible at much
shorter time scales than in the linear case.

\begin{figure}
\centerline{\includegraphics[width=\figwl\textwidth,angle=-90]{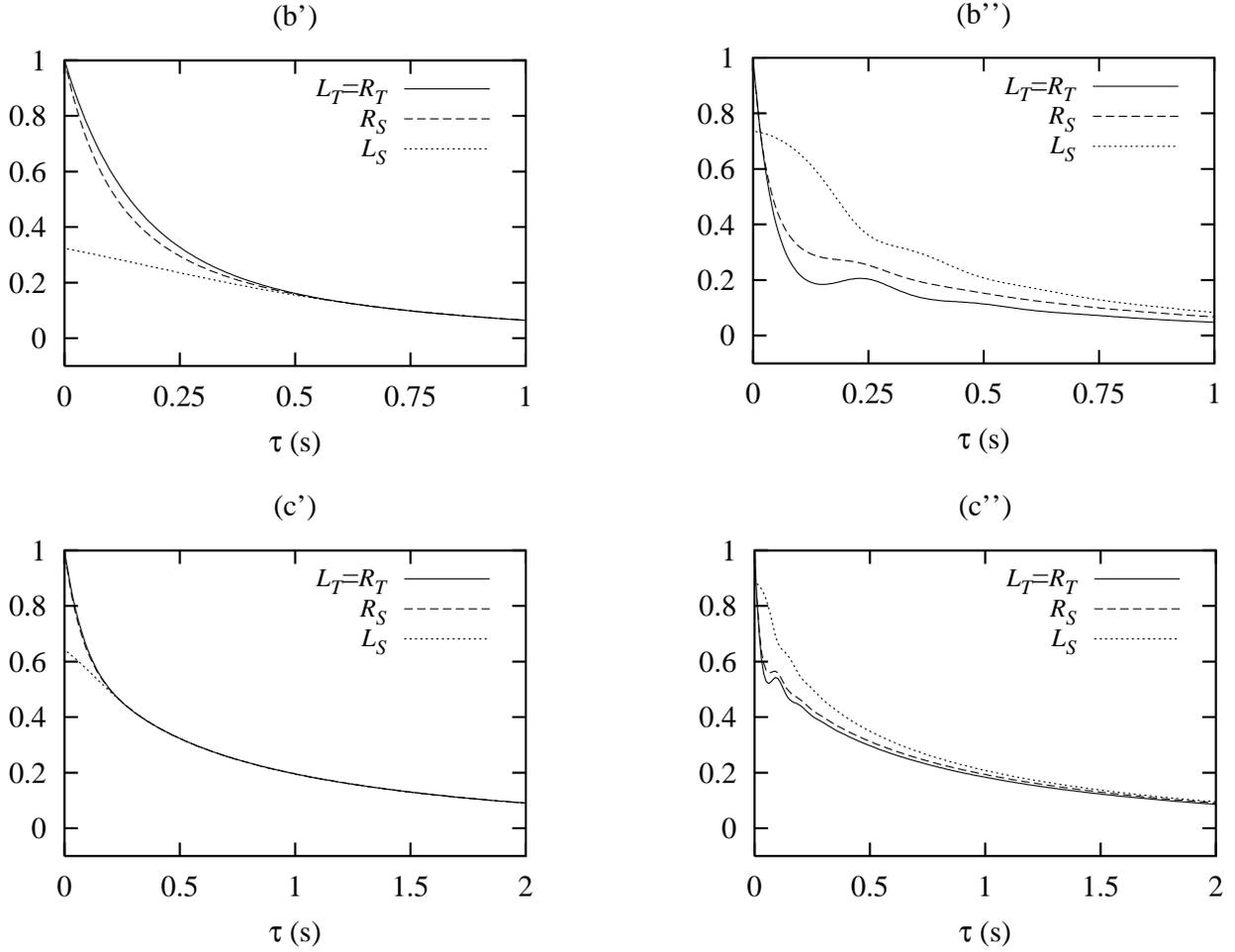}}
\caption{Integrated Fluctuation Theorems for the work for the circular
motion $L_T=R_T$ [cf. \Eq{STex0}], $L_S$ and $R_S$ versus
$\tau$ with (b') $\tau_r = 0.2\, \rm s$, $w = 4.8\, \rm s^{-1}$
(cf. Fig~\ref{fig1}b) at resonance $\Omega=1/\tau_r$, (b'') same
$\tau_r$ and $w$ but $\Omega=5/\tau_r$, (c') $\tau_r = 0.08\, \rm
s$, $w = 1.92\, \rm s^{-1}$ (cf. Fig.~\ref{fig1}c, $R_S$ is
indistinguishable from $L_T=R_T$) at resonance
$\Omega=1/\tau_r$, and (c'') same $\tau_r$ and $w$ but
$\Omega=5/\tau_r$. Note that in (b') and (c'), beyond a time
$\pi\tau_r$, all curves become indistinguishable.}
\label{fig3}
\end{figure}

In Fig.~\ref{fig3}, $L_T=R_T$, $L_S$ and $R_S$ from
Eqs.(\ref{STex0}-d) are shown for two cases, which for $\Omega=0$
become identical to the cases $(b)$ and $(c)$ in Fig.~\ref{fig1}. For
these cases we show what happens at their respective resonance points
$\Omega=1/\tau_r$ and what happens when $\Omega$ is larger,
$5/\tau_r$. In varying the frequency, we keep $w$ fixed, which
physically means we would have to adjust $r$ according to
\Eq{sigmacircle}.

{}From Fig.~\ref{fig3}b', we see that at the resonant point
$\Omega=1/\tau_r$, while $L_S$ could not be seen in the
$\Omega\approx0$ case (cf.~Fig.~\ref{fig1}b), it can now be seen, but
is still too small compared to $R_S$ to check the SSFT. If at
$\Omega=0$ the value of $L_S$ could be seen (Fig.~\ref{fig1}c), then
going to the resonance, improves the situation: in Fig.~\ref{fig3}c',
$L_S$ and $R_S$ approach each other after a far shorter time than in
Fig.~\ref{fig1}c, and become indistinguishable after $\tau=\pi\tau_r$,
because of the exponentially small difference between them as
discussed above. We remark that when the agreement is already good in
the $\Omega=0$ case (like in Fig.~\ref{fig1}d), going to the resonance
changes little.  Furthermore, we see from Fig.~\ref{fig3} that going
beyond the resonance ($\Omega=5/\tau_r$) will cause the curves to
deviate from each other again. We also note that above resonance, the
work fluctuations in the stationary state are larger than those in the
transient case, which is the opposite as for the linear motion. In
fact, the form \Eq{tvs} takes here is [using Eqs.~\eq{circleMT},
\eq{circleMS} and \eq{circleVS}]
\begin{equation}
    \frac{L_T(\tau)}{L_S(\tau)} 
    \stackrel{\tau\rightarrow\infty}{=}
\EXP{\left(\frac{1-\Omega^2\tau_r^2}{1+\Omega^2\tau_r^2}\right)
\frac{w\tau_r}{2}} 
\label{circlesvt},
\end{equation}
for $\tau\rightarrow \infty$. The exponent changes sign going from
$\Omega\tau_r<1$ to $\Omega\tau_r>1$.  What is happening physically is
that the system is driven so fast that it cannot relax within one
cycle. This increases the fluctuations $W_\tau$. On the basis of the
foregoing, we see that in order to demonstrate the ISSFT, it helps to
go to a resonant circular motion.

\section{Discussion}
\label{Discussion}

1. Inspired by the experiment of Wang {\em et al.\/}\cite{Wangetal02}
which showed that the work related ITFT holds when a small latex bead
is dragged linearly through a fluid by means of a laser-induced
harmonic force, a model of a Brownian particle in a harmonic potential
with an arbitrarily moving minimum was used to study both the work
related ITFT and ISSFT, under more general conditions.  From the
Langevin equation that describes the motion of the Brownian particle
in this simple model, everything can be explicitly calculated.  As
expected, the work related TFT holds for all time, as does its
integrated variant (ITFT).  The work related SSFT and its integrated
version (ISSFT) hold for sufficiently large times provided a
stationary state exists.

We also found a new relation between work related ISSFT and ITFT.  If
one looks at the ratios for the probability to find in a time $\tau$ a
negative vs. a positive work done on the system, in the transient
state ($L_T$) and in the stationary state ($L_S$), then $L_T/L_S$
approaches a constant (which is not $1$) as $\tau\rightarrow\infty$,
given by \Eq{tvs}.

2. We have not found many choices for the motion of the harmonic
potential ($\mathbf x^*_t$) for which a stationary state exists [in
the sense that the limits of \Eqs{MS}{VS} exist]. There is the linear
motion corresponding to the Wang {\em et al.\/} experiment, which has
been worked out in Sec.~\ref{sublinear}, and there is the possibility
of a circular motion treated in Sec.~\ref{subcircular}, as well as a
spiral motion, which is a trivial superposition of the previous
two. However, any motion which in the course of time approaches one of
these cases, will also reach a stationary state, corresponding to that
case.  In contrast to the simple motion of the harmonic potential
considered in Sec.~\ref{Applications}, allowing for an arbitrary
motion of the harmonic potential, may give rise to (arbitrarily)
different fluctuations in the transient and the stationary cases.

3. We note that on the basis of our explicit calculations, we are able
to explore under what conditions the work related ITFT and ISSFT might
be observable, which is relevant for the devise of future experiments.

For the ITFT, which holds for all time, observability is purely a
matter of how fast the quantity $L_T$ decays: if it decays too fast,
the ITFT will not be measurable. We showed that if one inserts values
taken from the Wang experiment\cite{Wangetal02} into the explicit
expression of $L_T$, the ITFT shows a clear signal which decays on the
order of a second, consistent with the fact that the ITFT could be
observed in that experiment. The relaxation time, i.e. $\tau_r$,
however is off; the relaxation time found in the experiment is of the
order of $1$ to $2$ seconds, whereas the value of the harmonic force
constant and the application of Stokes' Law give a relaxation time of
$0.5$ seconds. This could be due to boundary effects, local heating
due to the laser, deviations from the harmonic nature of the
laser-induced force.

The fact that the ITFT can be observed does not imply that the ISSFT
can be observed. In fact, problems with observability arise when we
insert values taken from the Wang experiment \cite{Wangetal02} into
the quantities for the ISSFT: the signal of $L_S$ is then too small
because the work done on the system in a relaxation time is too large.
There are a few ways to improve this situation, i.e., to make the
ISSFT observable in an experiment: first, one can take a smaller
particle, or second, make the velocity with which it moves through the
fluid smaller. Both methods reduce the work done in a relaxation time
$\tau_r$. If the diameter of the particle is reduced to about $16$
times smaller than the original one (which means about 400 nm across),
we see that the ISSFT can indeed be observed.

The circular motion that we investigated offers a third possibility to
improve the observability of the ISSFT: Under resonance conditions
($\Omega\tau_r=1$), the deviations from the ISSFT become exponentially
small after a time $\pi\tau_r$.

4. We end by giving some issues which are open for future
investigation. Some possible extensions of the theory could be the
following. The theory developed here is for the overdamped case
only. One might wonder if there is ever any practical need to consider
the situation where the damping is not so large. That would mean the
theory would start with a Langevin equation of motion for $\mathbf x$
and $\mathbf v$, or a Kramers equation.  We have carried out such
calculations for the case of a linear motion of the harmonic potential
and found the same results as reported here.  In the case of the
circular motion, this would also allow a more precise discussion of
the role of the centrifugal force (due to $\Omega$), which a rough
estimate limits to $\Omega\ll \alpha/m$ (this is of the order of $10^5
\rm Hz$ for parameters taken from the Wang {\em et al.\/}
experiment). One could also consider the case of anharmonic rather
than harmonic potentials.  On a practical level, is seems plausible
that the theory could be applied to other systems, such as linear
electrical circuits and could be particularly relevant to
nano-technology.

On a more fundamental level, one could ask what the precise relation
is of the work fluctuation theorem (for $\beta W^{tot}_\tau$)
discussed here and the usual entropy production theorems (for $\beta
W^{Br}_\tau$) for dynamical and stochastic
systems\cite{GallavottiCohen95a,GallavottiCohen95b,SearlesEvans00,Kurchan98,LebowitzSpohn99}.
While the work fluctuation theorems hold for all classes of systems
considered so far, this appears not to be the case for the entropy
production theorems.  In a future publication, we intend to discuss
this question in detail, since the theory needed for this deviates too
much from the present one to be included it in this paper. We can
state, however, that for the models considered here, such a theory
indicates that while an SSFT for the entropy production, i.e., for
$\beta W^{Br}_\tau$, appears to hold for long times as usual, the TFT
for the entropy production seems to hold {\em for long times only} as
well\cite{VanZonCohen03}, and not as an identity for all times, as
would be expected\cite{CohenGallavotti99}.

This work has been supported by the U.S. Department of Energy, under
grant No. DE-FG-02-88-ER13847.


\begin{thebibliography}{10}
\bibitem{footnote1}
Physically, it is not immediately clear that \Eq{WS} expresses any kind of
  work, since the force $\mathbf F$ is exerted on the particle at position
  $\mathbf x_t$, while $\mathbf v^*_t$ refers to the motion of the center of
  the harmonic potential. However, the force $\mathbf F$ on the particle
  implies a reaction force $-\mathbf F$ on the harmonic potential (laser),
  which must be overcome by an external force $\mathbf F$ exerted on the laser.
  Therefore, the force in \Eq{WS} is also the external force exerted on the
  laser and $W^{tot}_\tau$ is indeed the work performed on the system, e.g. by
  the experimentalist.
\bibitem{footnote2}
Mazonka and Jarzynski\cite{MazonkaJarzynski99} were aware of the difference
  between the entropy production and the total work done, although they used a
  different wording.
\end{thebibliography}
\end{document}